\documentclass[amsmath,amssymb,aps,pra,reprint,showpacs]{revtex4-1}
\usepackage{graphicx}
\usepackage{dcolumn}
\usepackage{color}

\def\R{\mathbb{R}}

\begin{document}
\title{Invisible waveguides on metal plates for plasmonic analogues of electromagnetic wormholes}
\author{Muamer Kadic}
 \affiliation{Aix-Marseille Universit\'e, CNRS, Centrale Marseille, Institut Fresnel, \\Campus universitaire de Saint-J\'er\^ome, 13013
Marseille, France}
\author{Guillaume Dupont}
 \affiliation{Aix-Marseille Universit\'e, CNRS, Centrale Marseille, Institut Fresnel, \\Campus universitaire de Saint-J\'er\^ome, 13013
Marseille, France}
\author{Stefan Enoch}
 \affiliation{Aix-Marseille Universit\'e, CNRS, Centrale Marseille, Institut Fresnel, \\Campus universitaire de Saint-J\'er\^ome, 13013
Marseille, France}
\author{Sebastien Guenneau}
\affiliation{Aix-Marseille Universit\'e, CNRS, Centrale Marseille, Institut Fresnel, \\Campus universitaire de Saint-J\'er\^ome, 13013
Marseille, France}
\begin{abstract}
We introduce two types of toroidal metamaterials which are invisible to surface plasmon
polaritons (SPPs) propagating on a metal surface. The former is a
toroidal handlebody bridging remote holes on the metal surface:
It works as a kind of plasmonic counterpart of electromagnetic
wormholes.
The latter is a toroidal ring lying on the metal surface: This bridges two disconnected metal
surfaces i.e. It connects a thin metal cylinder to a flat metal surface with a hole.
Full-wave numerical simulations demonstrate that an electromagnetic field propagating
inside these metamaterials does not disturb the propagation of SPPs at the metal surface.
A multilayered design of these devices is proposed, based on effective medium theory
for a set of reduced parameters: The former plasmonic analogue of electromagnetic wormhole requires homogeneous isotropic
magnetic layers, while the latter merely requires dielectric layers.
\end{abstract}

\maketitle
\section{Introduction}
Eight years ago, two groups of physicists \cite{pendry,leonhardt06} unveiled independent
paths towards electromagnetic invisibility. The transformational optics proposal by Pendry et al. leads to singular tensors on the frontier of the invisibility region \cite{greenleafa,greenleafb} that require an extreme electromagnetic response achieved upon resonance of split ring resonators \cite{pendryexp}. Various extensions including the blowup of a segment instead of a point \cite{jiang,wormhole1} and the stereographic projection of a virtual hyper-sphere in a four dimensional space \cite{tyc} have been studied. The conformal optics proposal by Leonhardt's grouping \cite{leonhardt06} leads to spatially varying, but bounded, scalar permittivity and permeability. However, the mathematical tools of complex analysis thus far constrain the invisibility design to two-dimensions.
Some recent advances in quasi-conformal optics \cite{lipendry,smithcarpet,zhangcarpet1,gabrielli-carpet} also found some applications in the control of surface plasmon polaritons (SPPs): Transformational plasmonics \cite{smol,spp1,zheludev,spp2,spp3,spp4,spp5,kadic2012}. Harnessing SPPs in order to deliver coupling between surface electrons on a structured metallic plate and incident light is a core topic in plasmonics \cite{bill,ebbesen,science2004,javier}, and plasmonic resonances underpins invisibility relying upon devices such as  out-of-phase polarizability shells with low refractive index \cite{engheta}, core-shell anomalous resonances \cite{milton2}, or concentric rings of point scatterers \cite{baumeier}.
However, the field of transformational plasmonics has a broader range of applicability as it is fuelled by analogies with Einstein's general relativity such as electromagnetic wave propagation in inhomogeneous media and particle/light motion in gravitational potentials. For example, the plasmonic Eaton lens proposed by Zhang's team \cite{spp5} is reminiscent of an optical black hole \cite{blackhole4,blackhole1,blackhole2}, which can trap and absorb electromagnetic waves coming from all directions.
In the present article, we adapt the design of transformation based wormholes to the area of surface plasmon polaritons (SPPs). In physics and fiction, a wormhole is portrayed as a shortcut through spacetime. Building upon the recent proposal of electromagnetic wormholes by Greenleaf et al. \cite{wormhole1}, it is enough to consider it as a topological feature of space. Our main contribution is an explicit design of a toroidal handlebody to control SPPs propagating at a metal surface with two holes: The main ingredient in the recipe of our plasmonic analogue of electromagnetic wormhole is to blow up a curve, rather than a point as is used in a typical three-dimensional invisibility cloak.  We further numerically demonstrate the validity of our theoretical approach with three-dimensional finite element computations for SPPs propagating in a toroidal heterogeneous anisotropic wormhole (we later abuse of the word wormhole to refere to our toroidal cloak). We finally derive some reduced set of parameters allowing for the design of a multi-layered toroidal tunnel consisting of an alternation of isotropic homogeneous layers approximating the ideal cloaking device in the homogenization limit. This brings our plasmonic analogues of electromagnetic wormholes a step closer to an experimental setup. Potential applications in plasmonics range from invisible plasmonic waveguides (which could be useful in making measurements of electromagnetic fields without disturbing them, or as new types of endoscopes in medical applications), to hard discs for optical computers (the latter requires a tilted version of the wormhole which lies on a metal surface, which is also
discussed). Other applications can be also envisaged thanks to the unprecedented control of SPPs through transformational plasmonics.

\section{Description of toroidal handlebodies on metal plates}
In this article, we introduce two types of plasmonic analogues of electromagnetic wormholes.
We start by describing the mathematical construction of
a magnetic SPP analogue to electromagnetic wormhole, which involves an electromagnetic toroidal cloak thereafter called
handlebody and two holes on a metal plate. Such a wormhole can be implemented for
electromagnetic fields by deriving the required
tensors of permittivity and permeability for a toroidal region
of $\R^3$ (the invisible tunnel) connecting two regions of
a metal surface, using the tools of transformational plasmonics.

\begin{figure}[ht!]
\centerline{
\hspace{0cm}\includegraphics[width=9cm,angle=0]{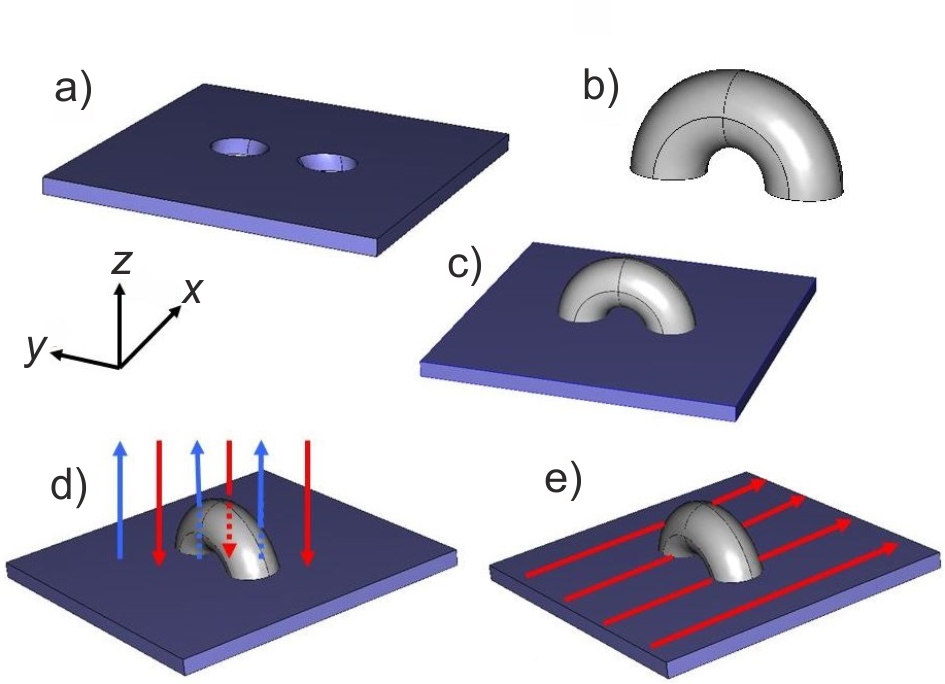}}
\vspace{0cm} \caption{(Color online) Main ingredients of plasmonic analogue of electromagnetic wormhole I:
diagrammatic view of the metal plate with two remote holes, $M=\R^2\setminus(D_1\cup D_2)$ (a), and the toroidal handlebody $T=\partial M\times [0,L]$ bridging the holes (b);  These form a Plasmonic analogue of electromagnetic wormhole $W=M\cup T$ (c);
Diagrammatic view of streamlines for a normally incident plane wave on the holey metal plate with handlebody $W=M\cup T$ (d) and for a surface plasmon polariton (SPP) propagating on the holey plate $M$
along the toroidal axis i.e. $x$-direction (e). The handlebody $T$ is invisible to any electromagnetic field (including SPPs).} \label{fig1-2}
\end{figure}

\begin{figure}[ht!]
\centerline{
\hspace{0cm}\includegraphics[width=9cm,angle=0]{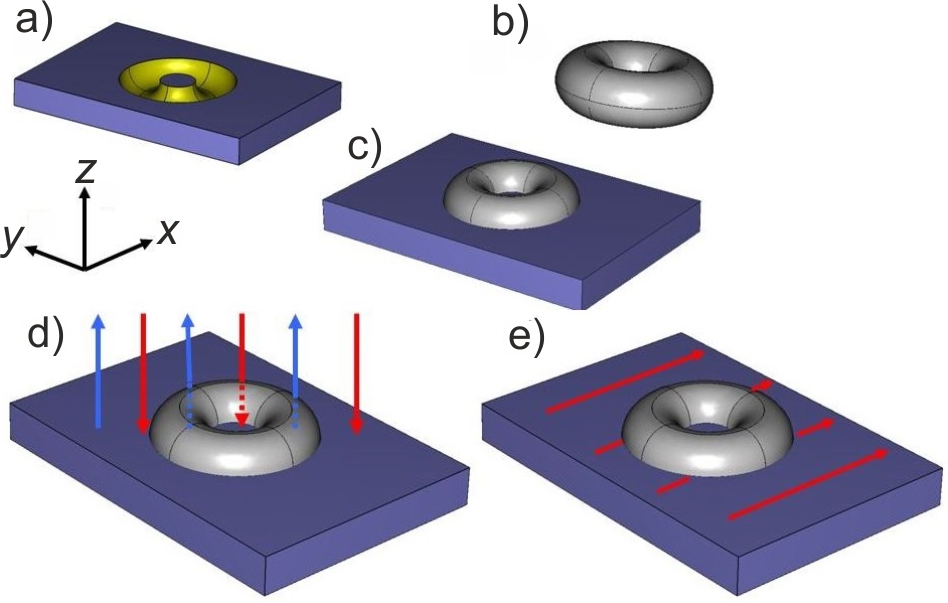}}
\vspace{0cm} \caption{(Color online) Main ingredients of plasmonic analogue of electromagnetic wormhole II:
diagrammatic view of the metal plate with a large hole (manifold $M_1=\R^2\setminus D_1$) with a disconnected circular plate in its center (manifold $D_2$) (a) with the gold region depicting the region to be occupied by the toroidal handlebody $T_1$ (b). This results in a single manifold $W_1=M_1\cup T_1$, as shown in (c); Diagrammatic view of streamlines for a normally incident plane wave on the holey metal plate $M_1$ with toroidal handlebody $T_1$ (d) and for a SPP propagating along the $x$-direction i.e. orthogonal to the toroidal axis ($z$-direction). Note that SPPs do propagate
inside the center disc (this circular plate is glued to the infinite holey plate).
} \label{fig3-4}
\end{figure}

The main ingredients of our wormhole construction are as follows:
We start by making two identical holes in a metal plate (see Fig.\ref{fig1-2}(a)), for instance
two discs $D_1$ and $D_2$ separated by some positive distance
on a plane.
We denote by $M$ the region so obtained:
$M=\R^2\setminus (D_1 \cup D_2)$. Topologically, $M$ is a two-dimensional manifold
with boundary, the boundary of $M$ being
$\partial M=\partial D_1\cup\partial D_2$. We note that $\partial M$ is
the disjoint union of two discs on the plane.

The second component of the SPP analogue to electromagnetic wormhole $W$ is a curved toroidal cylinder
(see Fig.\ref{fig1-2}(b)),
$T= \partial M \times [0;L]$, where $L$ denotes the arc-length which connects
points of circle $\partial D_1$ to points of circle $\partial D_2$.
As the boundaries of $M$ and $T$ are topologically the same ($\partial M=\partial T=\partial D_1\cup\partial D_2$),
we can glue these boundaries together. The resulting domain $W$ no longer lies
on the metal surface $\R^2$, but rather has the topology of Euclidian space $\R^3$
with a three-dimensional handle attached to it, see Fig.\ref{fig1-2}(c). 
W is a two-dimensional space with a special topology which is embedded in the Euclidean physical 3-dimensional space.
We show in Fig.\ref{fig1-2}(d) a diagrammatic view of streamlines of a planewave incident upon W (the ring-like handlebody is invisible, hence W reflects light like a flat metal surface) and ray trajectory of SSP propagating on W (there is no perturbation of the SPP trajectory induced by the holes and the ring-like handlebody), see Fig.\ref{fig1-2}(e).

Regarding the construction of the dielectric SPP analogue to electromagnetic wormhole, see Fig. \ref{fig3-4},
the previous construction repeats mutatis mutandis with the noticeable
difference that the manifold $M$ should be replaced by a manifold $M_1$
with a single hole: $M_1=\R^2\setminus D_1$. Moreover, the disc $D_2$ is now a piece
of metal located inside the hole $D_1$: $D_2\subset D_1$, see Fig. \ref{fig3-4}(a).
We fill in the hole with a toroidal cloak, see Fig. \ref{fig3-4}(b), which results in the
dielectric type of SPP analogue of electromagnetic wormhole, see Fig. \ref{fig3-4}(c).
We show in Fig.\ref{fig3-4}(a) a diagrammatic view of streamlines of a planewave incident upon $W_1$ (the handlebody
is invisible, hence $W_1$ reflects light like a flat metal surface) and ray trajectory of SSP propagating on $W_1$ (there is no
perturbation of the SPP trajectory by the holes and the handlebody), see Fig.\ref{fig3-4}(b). 

\section{Transformation plasmonics for the design of plasmonic analogues of electromagnetic wormholes}

\begin{figure}[ht!]
\centerline{
\hspace{0cm}\includegraphics[width=8.5cm,angle=0]{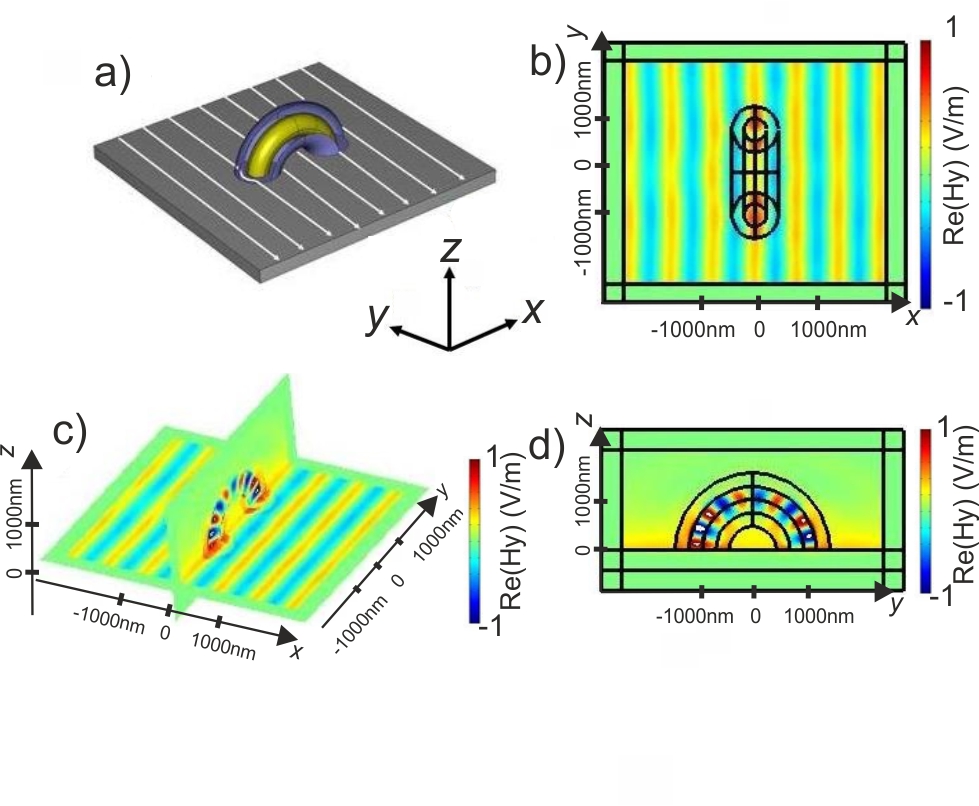}}
\vspace{0cm} \caption{(Color online) Principle of the undetectable
tunnel bridging two distant regions on a metal surface.
The electromagnetic waveguide is shown
in yellow and the coating in blue.
It is coated with a transformed dielectric plasmonic analogue of electromagnetic wormhole with reduced parameters: in principle, the handlebody is acting as a waveguide for electromagnetic waves launched from the metal dielectric interface where we set ${\bf H}=(H_{x2},0,0)\exp(-i\omega t)$ with $H_{x2}=\exp(-i\sqrt{2} z)$ (a); (b) Two dimensional plot (view from above) of the
real part of the magnetic field; (c) Three-dimensional plot validating the guiding and invisibility properties; (d) Two-dimensional plot of the real part of the magnetic field in the vertical plane showing the inner structure of the wormhole with a dielectric in the middle region which is surrounded by two regions of transformed medium. Note that the interfaces between the regions consist of a thin, conducting layer of thickness $70$ nanometers. These computations are for a wavelength of 700 nanometers for two remote circular holes of diameter $2a=533$ nanometers with a center to center spacing of $2R=1000$ nanometers.
} \label{fig5}
\end{figure}

\begin{figure}[ht!]
\centerline{
\hspace{0cm}\includegraphics[width=8.5cm,angle=0]{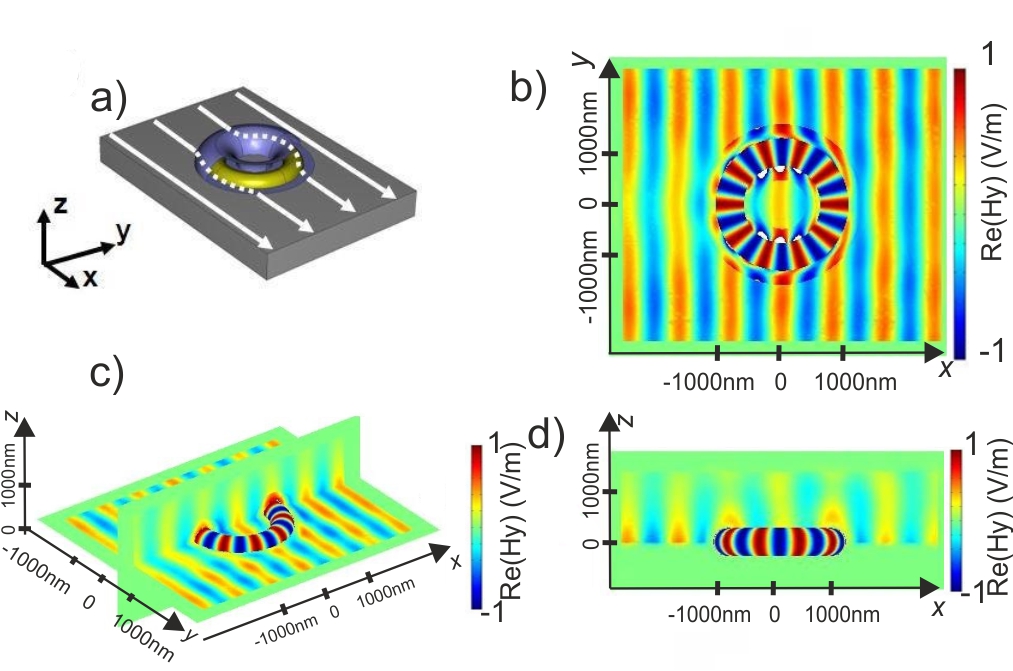}}
\vspace{0cm} \caption{(Color online) (a) Principle of the undetectable toroidal ring
bridging two disconnected regions i.e. a metal cylinder and a metal surface.
The electromagnetic waveguide is shown
in yellow and the coating in blue. Ray trajectories are drawn for illustrative purpose only.
Full wave simulations validate the theoretical proposal: (b) View from top; (c) 3D plot; (d) Side view;
Here, all plots are for the real part of the magnetic field and we set ${\bf H}=(0,0,H_{z2})\exp(-i\omega t)$ with $H_{x2}=\exp(-i\sqrt{2} z)$ on the waveguide cross-section in the vertical plane $y=0$.
} \label{fig6}
\end{figure}

We now wish to apply tools of transformational plasmonics
to design a device in $\R^3$ which controls the
propagation of SPPs in the same way as
the presence of the handle $T$ in the wormhole manifold
$W$. On $W$ we shall use the Riemannian metric that is the
Euclidian metric on $M$ and the product metric on $T$.
We emphasize that we are not actually tearing
and gluing plasmonic space, but instead prescribing
a metamaterial which makes the SPPs propagating on the
metal plate (see Fig. \ref{fig5} and \ref{fig6}) behave as if they
were propagating on the wormhole manifold $W$. In other
words, adopting the viewpoint of a SPP, it appears that
the topology of plasmonic space has been changed.

We first consider a surface plasmon polariton (SPP) propagating in the
positive $x$ direction at the interface $z=0$ between a metal surface ($z<0$)
and air ($z>0$), see Fig. \ref{fig5} and \ref{fig6}. If we choose the magnetic field ${\bf H}$ as the
unknown, it takes the following form:
\begin{equation}
\left\{
\begin{array}{ll}
{\bf H}_2 &= (0,H_{y_2},0) \exp \{i(k_{x2}x-\omega t)-k_{z2}z\} \; , z>0 \; ,\\
{\bf H}_1 &= (0,H_{y_1},0) \exp \{i(k_{x1}x-\omega t)+k_{z1}z \} \;
, z<0 \; ,
\end{array}
\right. \label{sppview}
\end{equation}
with $\Re(k_{z1})$ and $\Re(k_{z1})$ strictly positive in order to
maintain evanescent fields above and below the interface $z=0$.
The amplitude of the incident field has a Gaussian profile $H_{y_i}=e^{-\frac{y^2}{2\delta^2}}$ with $\delta=3\lambda$.

\subsection{Plasmonics at a metal plate-wormhole interface} 
Let us now consider two holes in the metal interface. It is clear that the propagation of the SPP
is affected by their presence, as illustrated in Fig. \ref{fig8}.
Our aim is to design an invisible handlebody through geometric surgery which will bridge the two holes
at the metal surface as shown in Fig. \ref{fig1-2}. For simplicity, we construct a device that has rotational
symmetry about a line in $\R^3$, and moreover we assume that the radii of $D_1$ and
$D_2$ are equal. We use toroidal 
coordinates $(r,u,v)$ corresponding to a point
$(x,y,z)=(r\cos u,(R+r\sin u)\sin v,(R+r\sin u)\cos v)$ in the Euclidean space $\R^3$, where $2R$ is the center-to-center
spacing between the discs $D_1$ and $D_2$, see Fig. \ref{fig13} in the Appendix.

Following the original proposal of electromagnetic wormholes by Greenleaf et al. \cite{wormhole1},
let us now consider the blowup of the centerline of the toroid which goes through the centers of $D_1$
and $D_2$, onto a toroidal coating using the transform $r'= a+ r(b-a)/b$, $u' = u$ and $v' = v$.
Here, $b$ and $a$ are the radii of the circles that form the outer and inner boundaries of the cloaking
region, respectively, in toroidal coordinates. Using the transformational plasmonics tools
\cite{spp3}, we obtain:

\begin{equation}
\begin{array}{ccc}
\varepsilon_{rr} & = \mu_{rr} =
\displaystyle{\frac{r-a}{r}\frac{(b-a)R+b(r-a) \sin u}
{(b-a)(R+r \sin u)}} \\
\varepsilon_{uu} &= \mu_{uu} =
\displaystyle{\frac{r}{r-a}\frac{(b-a)R+b(r-a) \sin u}{(b-a)(R+r \sin u)}} \\
\varepsilon_{vv} &= \mu_{vv} =
\displaystyle{\frac{b^2}{b-a}\frac{r-a}{r}\frac{R+r \sin u}{(b-a)R+b(r-a) \sin u}}
\end{array}\,.
\label{matorus}
\end{equation}

For a toroidal cloak, the angles $u$ and $v$ vary in the range $(-\pi,\pi)$ and $(0,2\pi)$. However,
for a plasmonic analogue of electromagnetic wormhole like Fig.\ref{fig1-2}, we only need the upper half part of a toroidal cloak,
that is $u$ varies between $0$ and $\pi$. If on the contrary we concentrate on the
toroidal cloak lying on the metal surface, see Fig.\ref{fig3-4}, we now need to tilt
the toroidal cloak by an angle of $\pi/2$ and further cut it in two halves along the z-axis,
so that it is now $v$ which varies from $0$ to $\pi$.

\section{Homogenization approach for a broadband multilayered
plasmonic analogue to electromagnetic wormhole}
\begin{figure}[ht!]
\centerline{
\hspace{0cm}\includegraphics[width=9cm,angle=0]{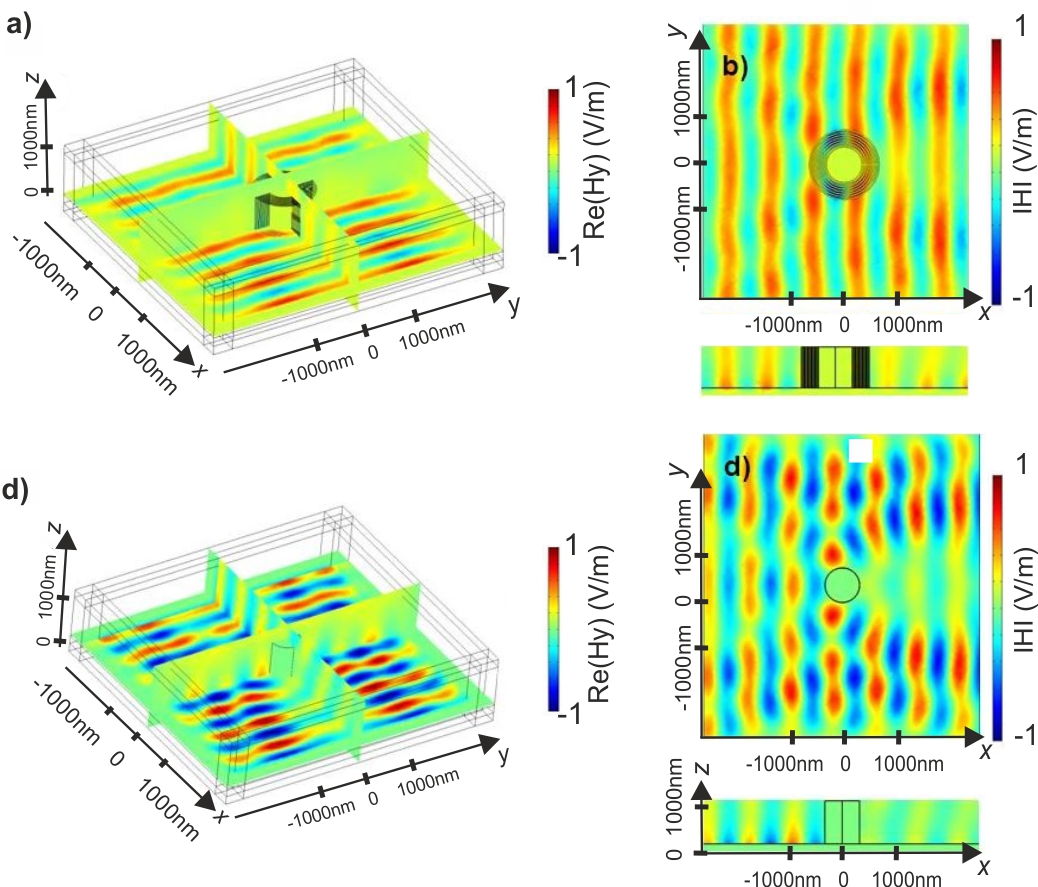}}
\vspace{0cm} \caption{(Color online) Structured cylindrical magnetic cloak: An SPP incident upon a multilayered cylindrical plasmonic cloak at 700 nanometers(a,b). The material parameters of the layered cloak can be found in [33], wherein a two-dimensional case was considered.; The same is plotted for comparison for an obstacle on its own (c,d). The reduced backward and foward scattering in (a,b) is noted.}
\label{fig7}
\end{figure}

We would like now to approximate the transformed medium associated with
the wormhole by some structured material. We opted for the homogenization
approach which should lead to a broadband metamaterial. Indeed, it is fairly
easy to extend the design of cylindrical multilayered cloaks originally proposed
by Huang et al. \cite{huang07} to surface plasmon polaritons, see Fig. \ref{fig7}.
There is however a further challenge in the present case: We need a multilayered
toroidal cloak. 

\begin{figure}[ht!]
\centerline{
\hspace{0cm}\includegraphics[width=9cm,angle=0]{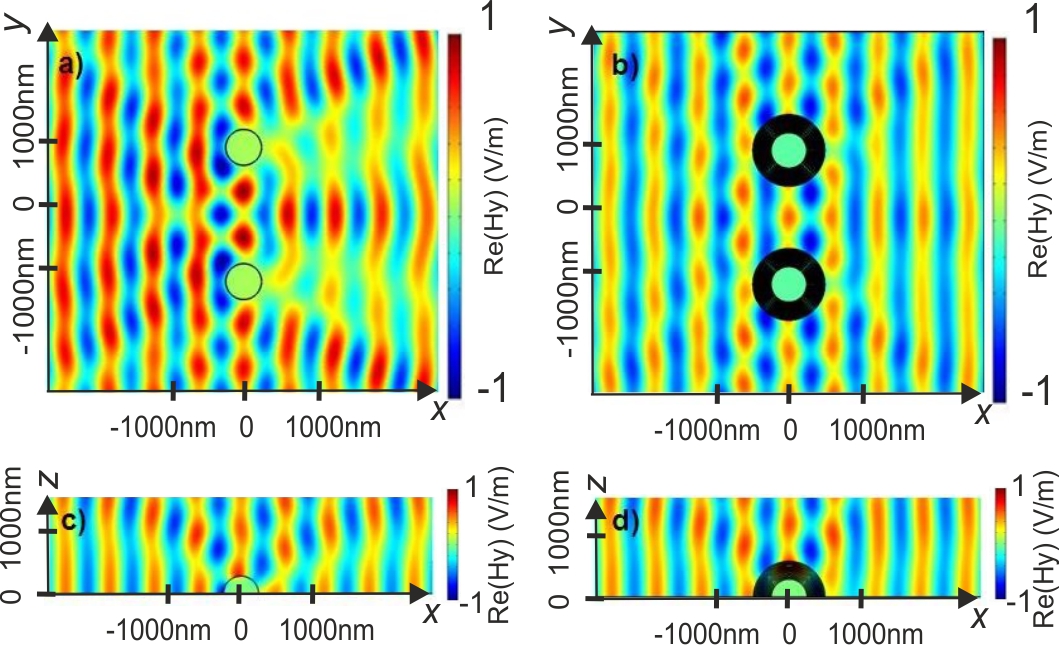}}
\vspace{0cm} \caption{(Color online) Structured cylindrical magnetic and dielectric cloaks: (a,b) Top view of the same configuration as in Fig. \ref{fig7} for two twin cylindrical magnetic cloaks (exemplifying the control of the SPP on the metal surface);
(c,d) Side view of an SPP incident from left upon a cylindrical metal bump on a metal surface on its own (c) and when it is surrounded by a cylindrical dielectric cloak with same parameters as in (b) but lying on the metal surface
(exemplifying the control of SPP above the metal surface);
The reduced scattering in (b) and (d) is noted.}
\label{fig8}
\end{figure}

\begin{figure}[ht!]
\centerline{
\hspace{0cm}\includegraphics[width=9cm,angle=0]{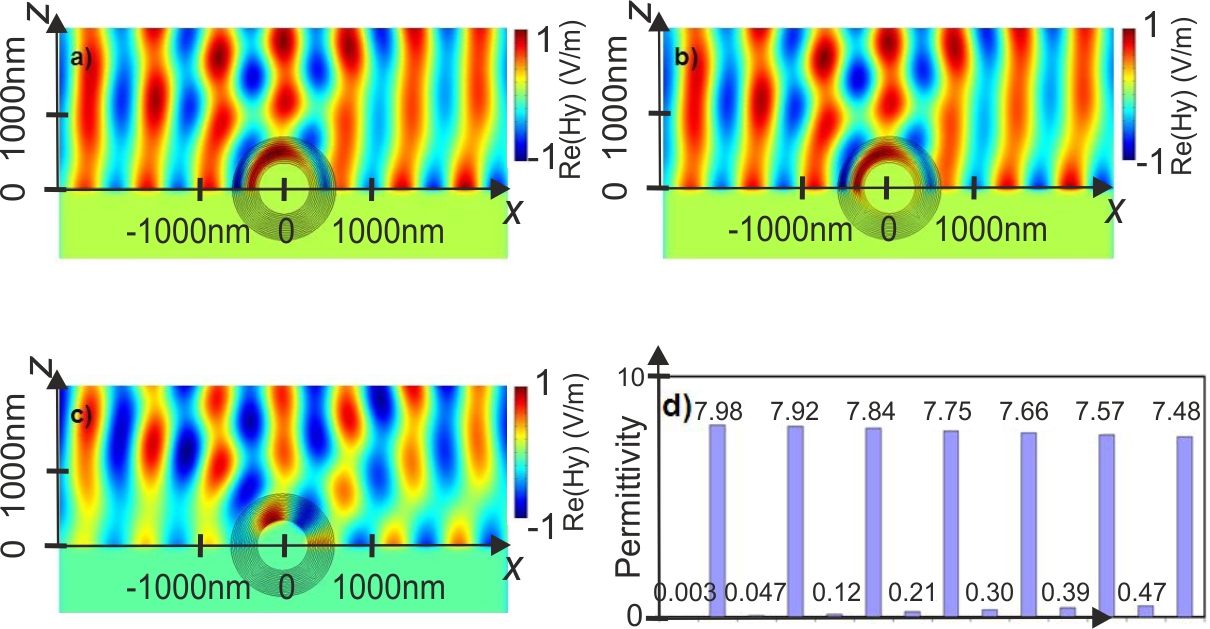}}
\vspace{0cm} \caption{(Color online) A SPP Gaussian beam with a waste of 2100
nanometers launched at 800 nanometers is incident upon the structured dielectric wormhole:
(a)  Side view of a multilayered dielectric wormhole surrounding a metal obstacle ; (b) Side view of the same multilayered
dielectric wormhole, when we complement it with concentric layers of metals in the metal plate;
(c) Side view for the metal obstacle on its own; (d) Values for epsilon inside the 14 homogeneous layers. The similar scattering
in (a) and (b) is noted.
} \label{fig9}
\end{figure}

For the construction of the wormhole of type I, we first consider two such cylindrical cloaks
located at the holes of the metal plate, see Fig. \ref{fig8}(a,b). It is reassuring to still
observe invisibility in that case. We then consider a structured cylindrical cloak lying on
the metal surface, and observe that the scattering of an SPP by a metal obstacle is
much reduced, see Fig. \ref{fig8}(c,d).
One question that might arrise is whether cloaking still works if we now bend the cylindrical cloaks
and whether the wormhole is broadband. We show
in Fig. \ref{fig9} that these results still hold at 800 nanometers for a structured tororidal ring
which shows that our broadband homogenization approach to this problem is
legitimate. Moreover, it is clear from  Fig. \ref{fig9} that there is no need to
structure the metal below the air-dielectric interface. This suggests a wormhole of type II
might consist only of homogeneous isotropic dielectric layers and might be therefore
easier to manufacture than the wormhole of type I.
However, the result in Fig. \ref{fig9} does not clearly show the effect of the curvature of the cloak.
In order to give a global overview of cloaking for the dielectric toroidal ring, we show
in Fig. \ref{fig10} that a line source emitting a concentric SPP
at the metal surface is not perturbed by the presence of a structured toroidal ring.
Importantly, all the previous results were obtained for visible light.
We are now ready to theoretically investigate
the design of structured toroidal cloaks via effective medium (homogenization) theory.

\subsection{Reduced parameters for a toroidal cloak}

In order to simplify the design of the wormhole, we proceed in a way similar to what was done to obtain
a reduced set of material parameters for cylindrical cloaks in \cite{naturecloak}.
Using the transformational plasmonics tools, we obtain: 
\begin{equation}
\begin{array}{ccc}
\varepsilon_{rr} & = \mu_{rr} =
\displaystyle{\frac{r-a}{r} f(r,u)} \\
\varepsilon_{uu} &= \mu_{uu} =
\displaystyle{\frac{r}{r-a}f(r,u)} \\
\varepsilon_{vv} &= \mu_{vv} =
\displaystyle{\frac{b^2}{{(b-a)}^2}\frac{r-a}{r \, f(r,u)}}
\end{array}
\end{equation}
where $f(r,u)=\frac{(b-a)R+b(r-a) \sin u}
{(b-a)(R+r \sin u)}$, which leads us to the set of reduced parameters
\begin{equation}
\begin{array}{ccc}
\varepsilon_{rr} &= \mu_{rr}& =
\displaystyle{{\left(\frac{r-a}{r}\right)}^2} \\
\varepsilon_{uu} &= \mu_{uu}& =
\displaystyle{1} \\
\varepsilon_{vv} &= \mu_{vv} &=
\displaystyle{\frac{b^2}{{(b-a)}^2}}
\end{array}
\end{equation}
that preserve the wave trajectories, but induce a slight impedance mismatch on the
wormhole ounter boundary.

For a toroidal cloak, both angles $u$ and $v$ vary in the range $(0,2\pi)$. However,
for a plasmonic analogue of electromagnetic wormhole, see Figure \ref{fig1-2}(a), we only need the upper half of a toroidal cloak,
that is $u$ varies between $0$ and $\pi$. If on the contrary we concentrate on the
toroidal cloak lying on the metal surface, see Fig. \ref{fig3-4}, we now need to tilt
the toroidal cloak by an angle of $\pi/2$ and further cut it in two halves along the z-axis,
so that it is now $v$ which varies from $0$ to $\pi$.

\subsection{Homogenized parameters for a toroidal cloak}

We note that if the toroidal cloak component of the wormhole consists of an alternation of two
homogeneous isotropic layers of thicknesses $d_A$ and $d_B$ and
permittivity $\varepsilon_A=\lambda_A^{-1}$, $\varepsilon_B=\lambda_B^{-1}$
and permeability $\mu_A=\lambda_A$ and $\mu_B=\lambda_B$, we have
\begin{equation}
\displaystyle{\frac{1}{\lambda_{rr}}}=\displaystyle{\frac{1}{1+\eta}\left(\frac{1}{\lambda_A}
+\frac{\eta}{\lambda_B}\right)} \; , \;
\lambda_{uu}=\lambda_{vv}=\displaystyle{\frac{\lambda_A+\eta \lambda_B}{1+\eta}}
\end{equation}
where $\eta=d_B/d_A$ is the ratio of thicknesses for layers $A$ and
$B$ and $d_A+d_B=1$.

\begin{figure}[ht!]
\centerline{
\hspace{0cm}\includegraphics[width=9cm,angle=0]{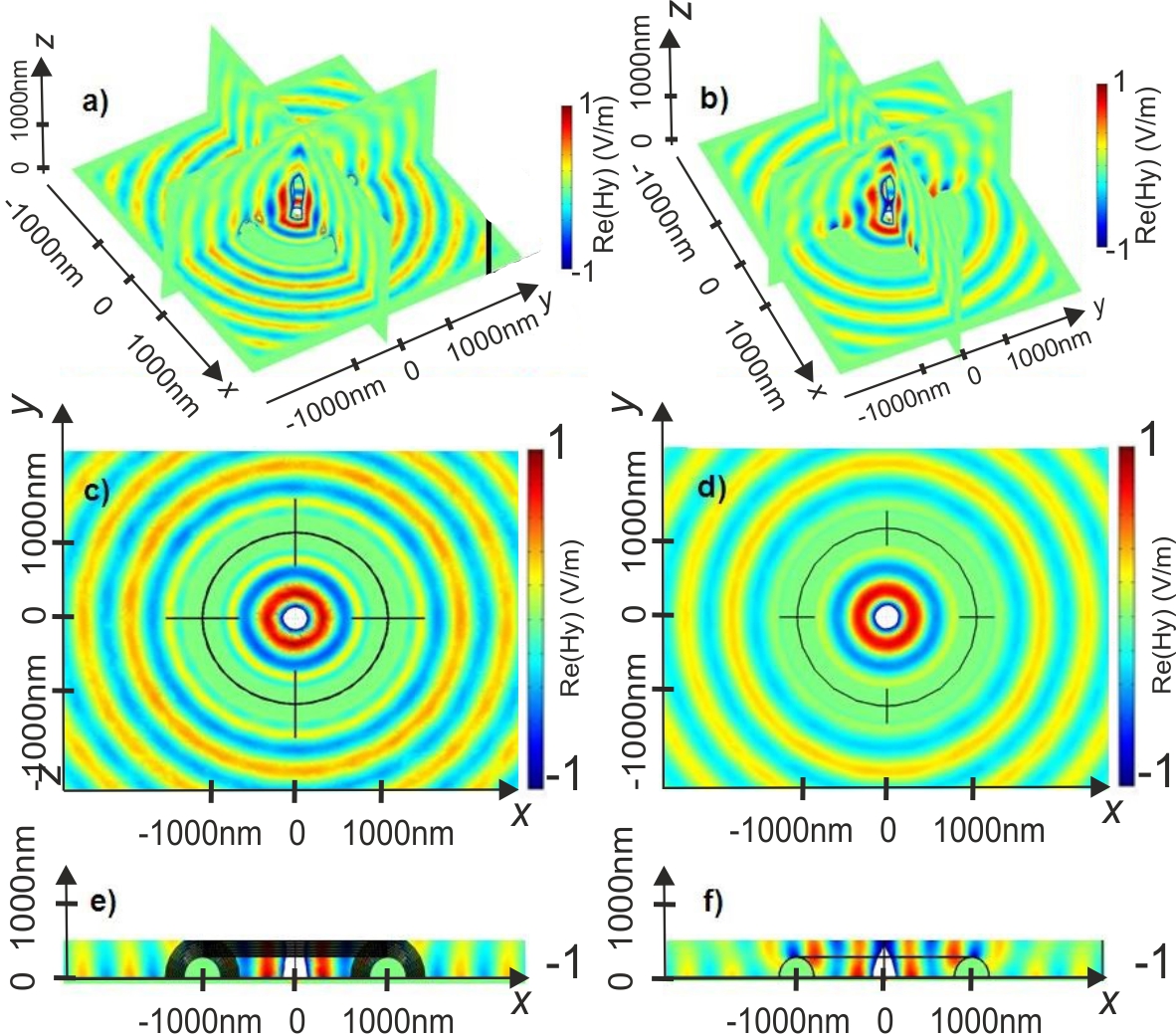}}
\vspace{0cm} \caption{(Color online) A point source generating a concentric SPP placed in the center of the system at a wavelength
of 700 nanometers is much less perturbed by the presence of the wormhole (as shown by the isotropic wave pattern
in the view from above and the sideview for the evanescent part of the SPP) than by the presence of an obstacle on its own.}
\label{fig10}
\end{figure}

\noindent We shall use the previous homogenized formulas in order to approximate the reduced parameters (4).

\begin{figure}[h]
\centerline{
\hspace{0cm}\includegraphics[width=8cm,angle=0]{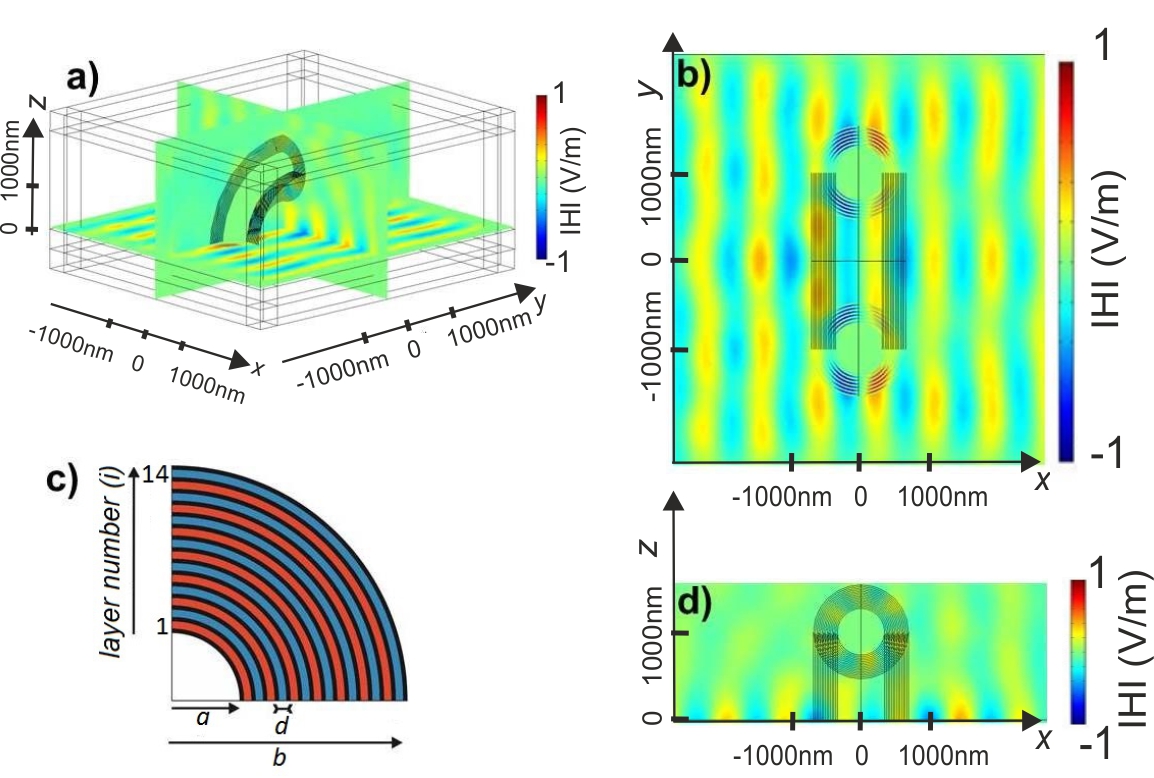}}
\vspace{0cm} \caption{(Color online) Structured magnetic wormhole: SPP Gaussian beam with a waste of $2100$ nanometers incident upon a multilayered cylindrical wormhole at 700 nanometers smoothly bent around a metal toroidal obstacle; The permeability in every layer (of identical thickness $60$nm) is given by $\mu_i$$=[$ $0,0032;$ $7,97;$ $0,0467;$ $7,91;$ $0,121;$ $7,84;$ $0,207;$ $7,75;$ $0,297;$ $7,66;$ $0,386;$ $7,57;$ $0,473;$ $7,48$ $]$, (from the inner to the outer layer);
(a) Three-dimensional plot of the real part of the magnetic field; (b) Top view; (c) Diagrammatic view of the device; (d)  Side view. The color scale has been normalized.} \label{fig11}
\end{figure}

\begin{figure}[h]
\centerline{
\hspace{0cm}\includegraphics[width=8.5cm,angle=0]{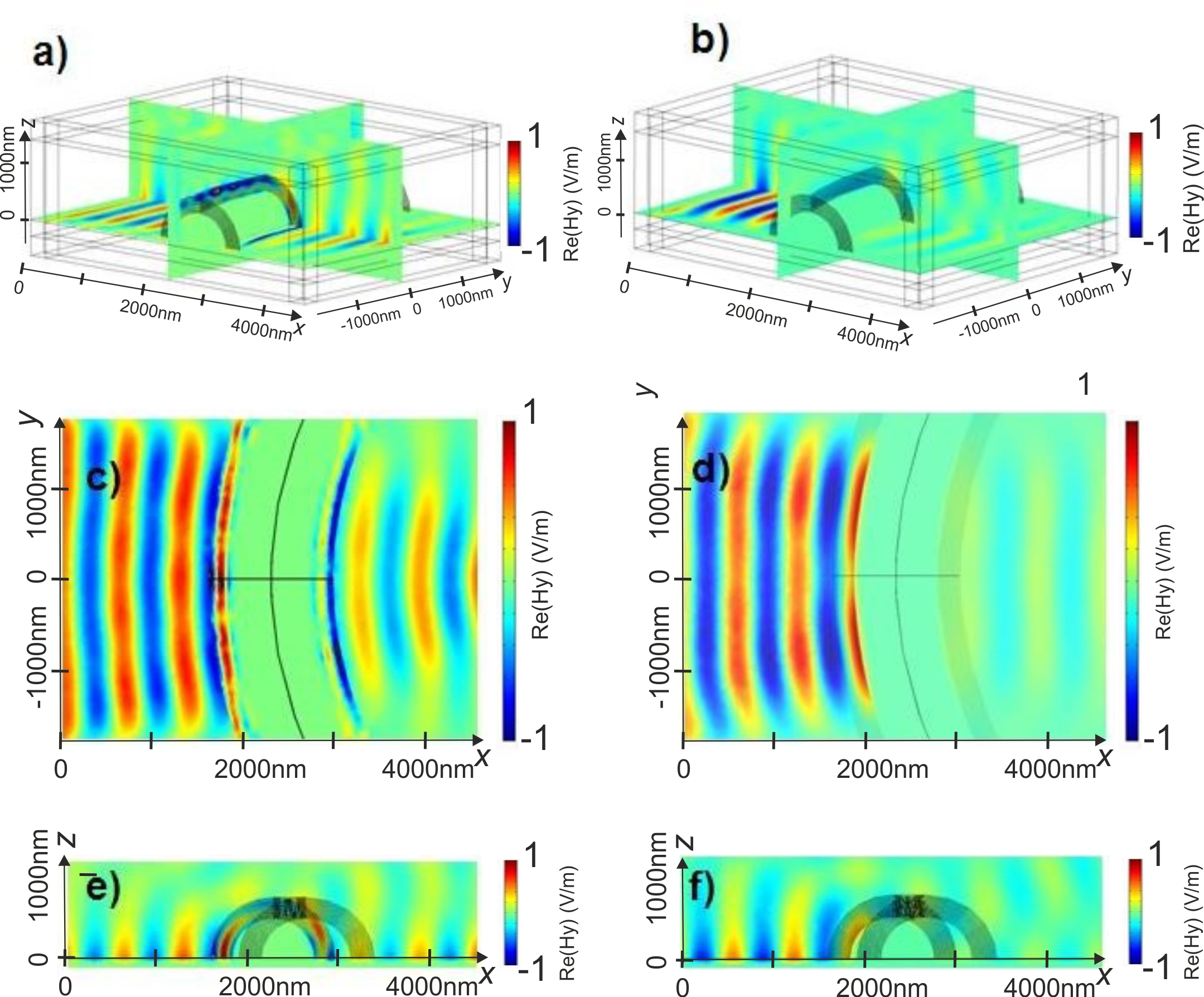}}
\vspace{0cm} \caption{(Color online) Structured dielectric wormhole:
SPP Gaussian beam with a waste of 2100 nanometers incident at 700 nanometers upon a metal toroidal obstacle dressed with the wormhole (left panel) and on its own (right panel). The Permittivity in every layer (of identical thickness $60$nm) is given by: $\varepsilon_i$$=[$ $0,0032;$ $7,97;$ $0,0467;$ $7,91;$ $0,121;$ $7,84;$ $0,207;$ $7,75;$ $0,297;$ $7,66;$ $0,386;$ $7,57;$ $0,473;$ $7,48$ $].$; (from the inner to the outer layer); 
(a,b) Three-dimensional plot; (c,d) View from above; (e,f) Side view. The color scale has been normalized.} \label{fig12}
\end{figure}

We report in Fig. \ref{fig11} some computations for a SPP incident at $700$ nanometers upon the structured magnetic wormhole consisting of an alternation of homogeneous magnetic layers specified in the figure caption, which could be achieved as in \cite{wood}. If we now tilt the wormhole by an angle $\pi/2$, a similar design holds with an alternation of dielectric homogeneous layers,
see Fig. \ref{fig12}. The performance of this dielectric SPP analogue to an electromagnetic wormhole is further ascertained by placing a plasmonic source in its center and observing
the unperturbed concentric wavefronts emanating from the source, see Fig. \ref{fig10}.

\section{Conclusion}
In conclusion, we have studied
analytically and numerically the extension of wormholes to
the domain of surface plasmon waves propagating at the interface
between metal and dielectric/air.
These waves obey the Maxwell equations at a flat interface and are
evanescent in the transverse direction, so that, the problem we have
treated is somewhat a two-dimensional plasmonic analogue of the electromagnetic wormhole designed by Greenleaf et al. \cite{wormhole1}: It is enough to consider the wormhole
as a manifold in a three dimensional Euclidean space for applications in plasmonics.
Nevertheless, our numerical computations based on the finite element
method take into account the three dimensional features of the
problem, such as plasmon polarization and jump of permittivity at
the interface between metal and metamaterial/air which are described
by permittivities of opposite sign.

Electromagnetic wormholes \cite{wormhole1} represent a fascinating
paradigm, but were initially thought of as an abstract
metamaterial bridging two spherical holes, thereby requiring a further
dimension for the invisible tunnel, and moreover no permittivity
and permeability tensors were derived from a specific structured design.
 We have transposed this idea to the area of surface plasmon polaritons, with two illustrative examples
of plasmonic analogues of electromagnetic wormholes (SPP analogues of volume electromagnetic effects described in \cite{wormhole1}):
An invisible handlebody and an invisible ring over metal surfaces. We further proposed 
multi-layered versions of these metamaterials which work
over a finite range of visible frequencies, hoping to foster experiments
in the emerging field of transformational plasmonics. Potential applications
might be in safer communications and intra-ship technologies.

\section*{Acknowledgements}
S.G. is thankful for a European funding through ERC Starting Grant ANAMORPHISM.

\onecolumngrid

\appendix

\section{Soft-Hard conditions}
In what follows, we numerically demonstrate that enforcing some {\it non-natural} conditions on the inner boundary, known as Soft-Hard Conditions:

\begin{equation}
{\bf e}_\theta.{\bf E}=0, \;\ {\bf e}_\theta.{\bf H}=0
\end{equation}
where ${\bf e}_\theta$ is the angular unit vector and $E$ and $H$ the local electric and magnetic field, leads to a farfield pattern very similar to Perfect Magnetic conditions.

\noindent For this, we extend the SH conditions to our wormhole as per: 
\begin{equation}
{\bf e}_w.{\bf E}=0, \;\ {\bf e}_w.{\bf H}=0
\label{sh1}
\end{equation}
where ${\bf e}_w$ is the angular unit vector in the toroidal basis $(r,w,v)$.

\begin{figure}[ht!]
\centerline{
\hspace{0cm}\includegraphics[width=7cm,angle=0]{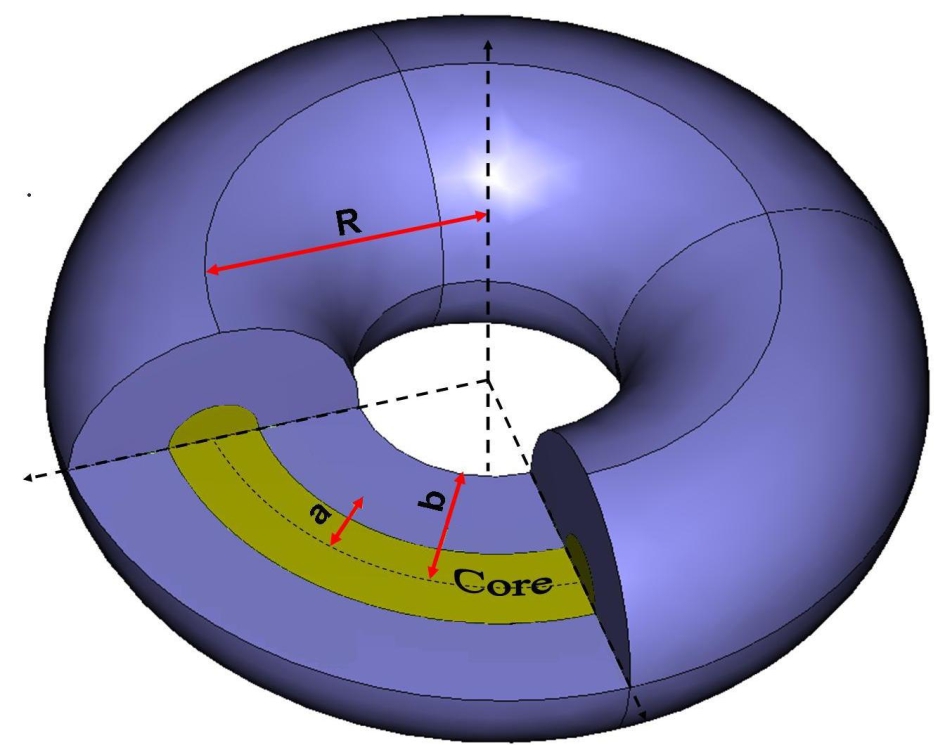}}
\vspace{0cm}
\caption{(Color online) Diagrammatic view of a toroidal wormhole: $R$ is the major radius of the torus, $a$ and $b$ are respectively the radii of the inner and outer boundaries
of the toroidal wormhole. The core (an isotropic dielectric medium shown in yellow color) corresponds to the invisible waveguide and the coating (the
transformed medium described by tensors of permittivity and permeability shown in blue color) allows the control of surrounding electromagnetic field,
e.g. SPPs.}
\label{fig13}
\end{figure}

\noindent The main difficulty is to impose such condition in a finite element algorithm, in our case Comsol Multiphysics. 
We start with the case of an $x-$ axis torus on a metal plate. 
The parametrisation is the following one: 
\begin{equation}
\left\{
\begin{array}{lll}
x &= r.\cos(w)\\
y &= (R+r. \sin(w)) \sin(v)\\
z &= (R+r. \sin(w))\cos(v)
\end{array}
\right.
\end{equation}

\begin{figure}[ht!]
\centerline{
\hspace{0cm}\includegraphics[width=15cm,angle=0]{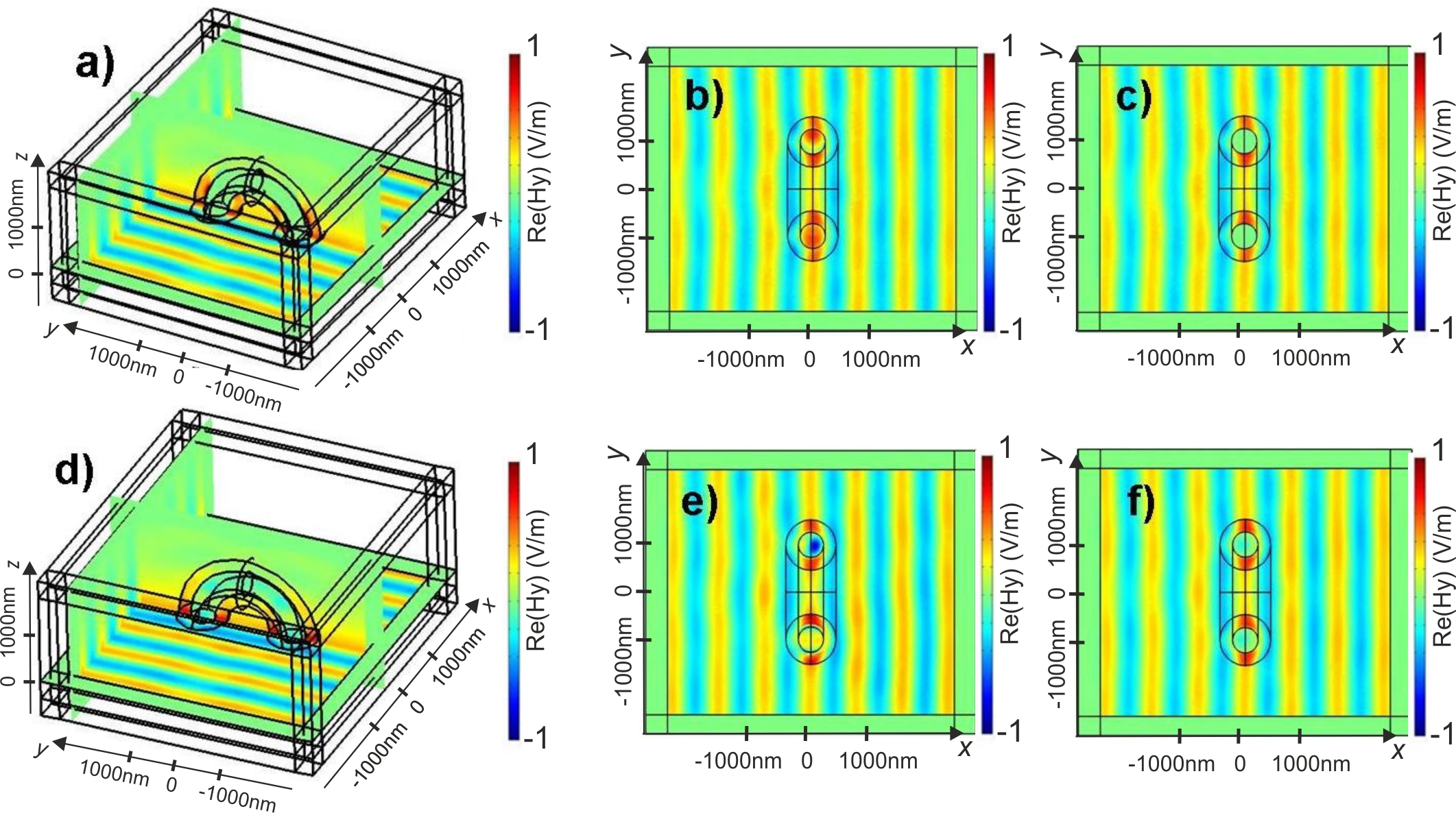}}
\vspace{0cm}
\caption{(Color online) Perfect magnetic conductor (PMC) versus Soft-Hard (SH) lining condition at $700$nm (The incident field is set as described in Eq. (1) of the accompanying letter with the field amplitude $H_{y_i}=1$) : (a) 3D Plot of the total magnetic field for PMC condition on the
inner boundary of the wormhole without propagating field inside the core;
(b) 2D Plot of the total magnetic field for PMC condition on the inner boundary of the wormhole with propagating field inside the core;
(c) 2D Plot of the total magnetic field for PMC condition on the inner boundary of the wormhole without propagating field inside the core;
(d) 3D Plot of the total magnetic field for SH lining condition on the
inner boundary of the wormhole without propagating field inside the core;
(e) 2D Plot of the total magnetic field for SH lining condition on the inner boundary of the wormhole with propagating field inside the core;
(f) 2D Plot of the total magnetic field for SH lining condition on the inner boundary of the wormhole without propagating field inside the core.}
\label{fig14}
\end{figure}

The vector ${\bf e}_w$ is given by:
\begin{equation}
\left\{
\begin{array}{lll}
x &= -r. \sin(w)\\
y &= r.\cos(w) \sin(v)\\
z &= r.\cos(w)\cos(v)
\end{array}
\right.
\end{equation}

\section{Finite element model and illustrative numerical examples}

We opted for Nedelec or edge finite elements, which naturally fulfill the tangential continuity of the electromagnetic field across interfaces. Put in a more mathematical way, mixed finite elements are part of a discrete algebraic-geometric-differential structure of finite element shape functions invented by H. Whitney \cite{w1} which assign degrees of freedom to simplices of a given mesh: nodes, edges, facets, tetrahedra. This structure, the so-called Whitney complex, closely matches a continuous structure made of four vector subspaces of L2 and of three differential operators grad, curl, div, which is known as the de Rham complex. This complex is called an exact sequence if the image of each operator domain of this structure is exactly the kernel of the next operator. Clearly, this statement depends upon the topological properties of the domains such as connectivity assumptions. This is why we choose the point of view of differential geometry in this article: It makes the numerical modeling much easier to handle. In this way, we indeed consider a simplicial mesh on a three-dimensional manifold W, that is, a set of tetrahedra which 2 by 2 have in common either a full facet, or a full edge, or a node (vertex), or nothing, and whose set union is W.

\begin{figure}[ht!]
\centerline{
\hspace{0cm}\includegraphics[width=9cm,angle=0]{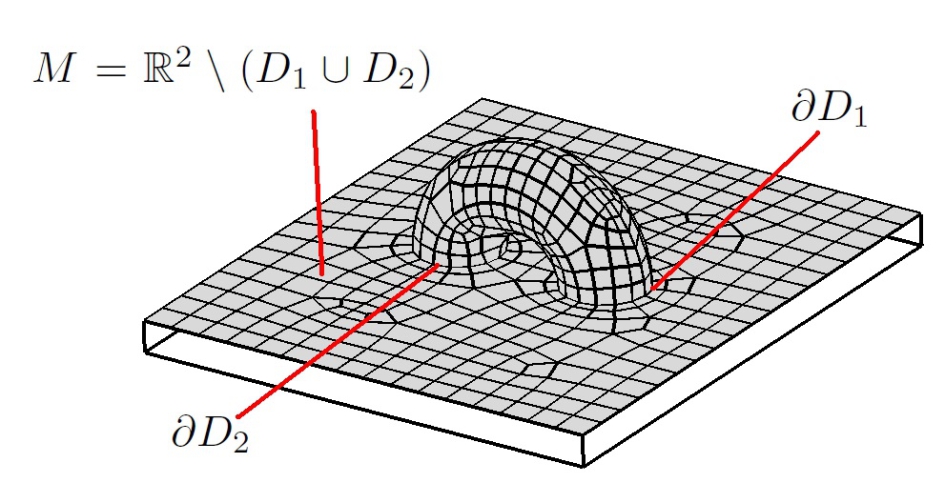}}
\vspace{0cm}
\caption{(Color online) Three-dimensional tetrahedral mesh for the two-dimensional manifold $W=M\cup T$, where $M=\R^2\setminus(D_1\cup D_2)$ for the plasmonic analogue of electromagnetic wormhole 1, where $D_1$ and $D_2$ are the remote holes (discs) in the metal plate.}
\label{fig15}
\end{figure}

In Comsol Multiphysics, we decided to solve the problem for the magnetic component ${\bf H}$ of the electromagnetic field:
\begin{equation}
\nabla\times\left(\underline{\underline\varepsilon}^{-1}\nabla\times{\bf H}\right)=\omega^2\varepsilon_0\mu_0\underline{\underline\mu}{\bf H}
\label{maxweak}
\end{equation}
where $\mu_0\varepsilon_0=c^{-2}$, with $c$ the speed of light in vacuum and $\underline{\underline\varepsilon}$ and $\underline{\underline\mu}$ are
tensors of relative permittivity and permeability proportional to the identity matrix outside the wormhole
and fully anisotropic and highly heterogeneous within the wormhole. Here, the unknow ${\bf H}={\bf H}_i+{\bf H}_d$, where ${\bf H}_i$ is the incident
field and ${\bf H}_d$ is the diffracted field which satisfies the usual outgoing wave conditions (to ensure existence and
uniqueness of the solution). The weak formulation associated with (\ref{maxweak} is discretised using
second order finite edge elements (or Whitney forms) which behave nicely under geometric
transforms (pull-back properties) \cite{andre}.

\noindent Since the above equation is taken in weak sense, it contains within it the transmission conditions across interfaces between different media.
In particular, the tangential continuity of the solution across the inner boundary of the wormhole reads as:
\begin{equation}
{\bf n}\cdot\left(\underline{\underline\varepsilon}^{-1}\nabla\times{\bf H}\right)=0 \; ,
\end{equation}
where ${\bf n}$ is the unit outward normal to the boundary.
 
\noindent The SH condition (\ref{sh1}) should therefore be reexpressed into:
\begin{equation}
{\bf e}_w.\left({\underline{\underline\varepsilon}}^{-1}\nabla \times {\bf H}\right)=0 \; , \; {\bf e}_w.{\bf H}=0.
\label{sh2}
\end{equation}

\noindent Such conditions can be enforced in COMSOL using the built-in constraint dialog box in boundary settings-Equation system,
see \cite{comsol}. However, other finite element packages may also be able to handle this model.

\noindent The permittivity and permeability in the wormhole are easily deduced through the relationship $\underline{\underline\varepsilon}=\underline{\underline{\mu}}={\bf T}$ \cite{andre},
where the entries of the transformation matrix ${\bf T}$ are given in the next section. The implementation of the SH lining
condition on the inner boundary follows from the conditions (\ref{sh2}).

\begin{figure}[ht!]
\centerline{
\hspace{0cm}\includegraphics[width=15cm,angle=0]{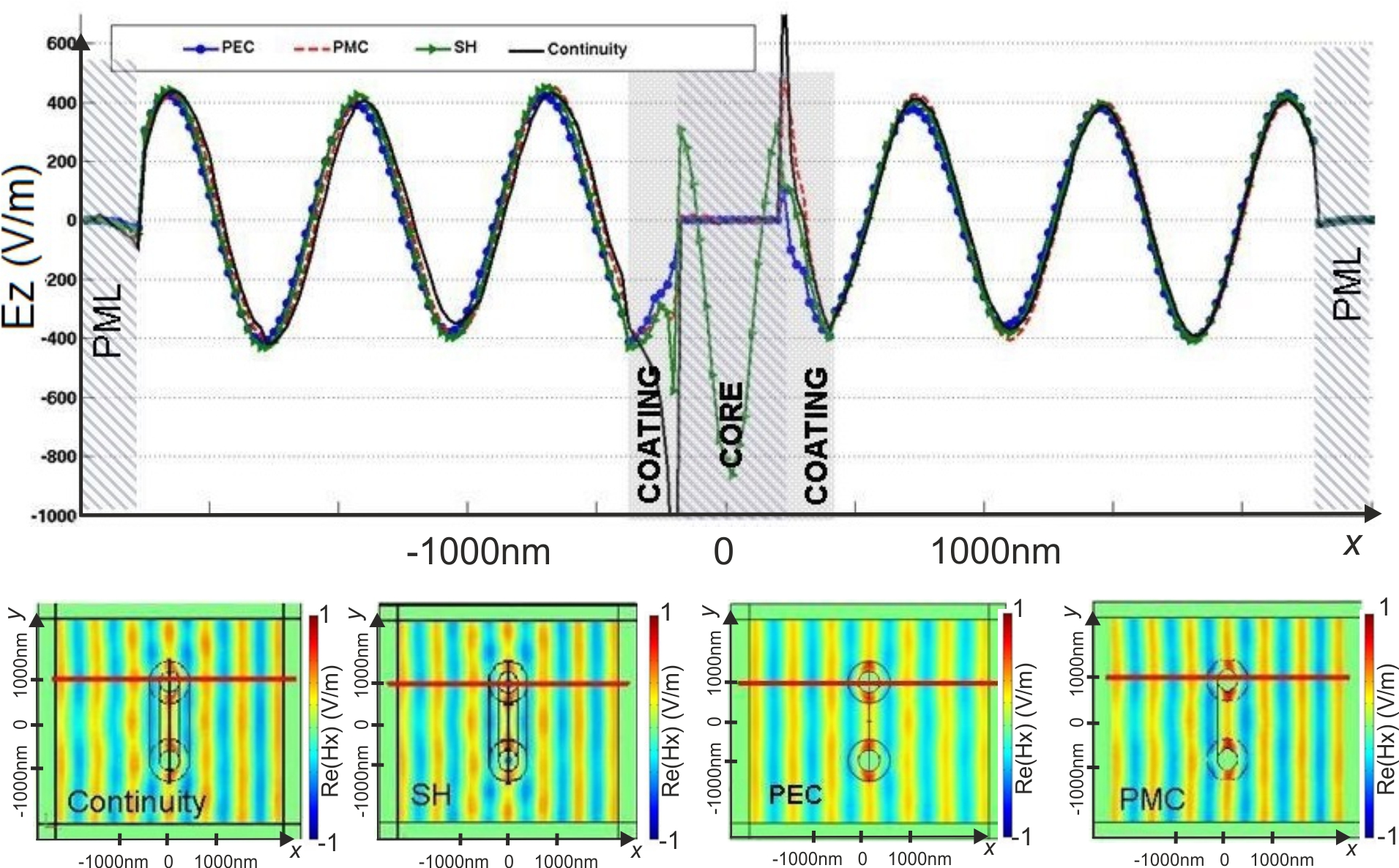}}
\vspace{0cm}
\caption{(Color online) Profile of $E_z$ in the plane $z=0$ along the red line shown in the inset, for three conditions on the inner boundary of the toroidal wormhole: Soft-Hard lining condition (dotted green curve), continuity condition (continuous black curve), Perfect Electric conductor (PEC in blue) and Perfect magnetic Conductor (PMC). We note for PMC condition, all that along this line both components $E_x$ and $E_y$ vanish for the SH condition). The $E_z$ component is not set to $0$ for SH condition.}
\label{fig16}
\end{figure}

\begin{figure}[ht!]
\centerline{
\hspace{0cm}\includegraphics[width=15cm,angle=0]{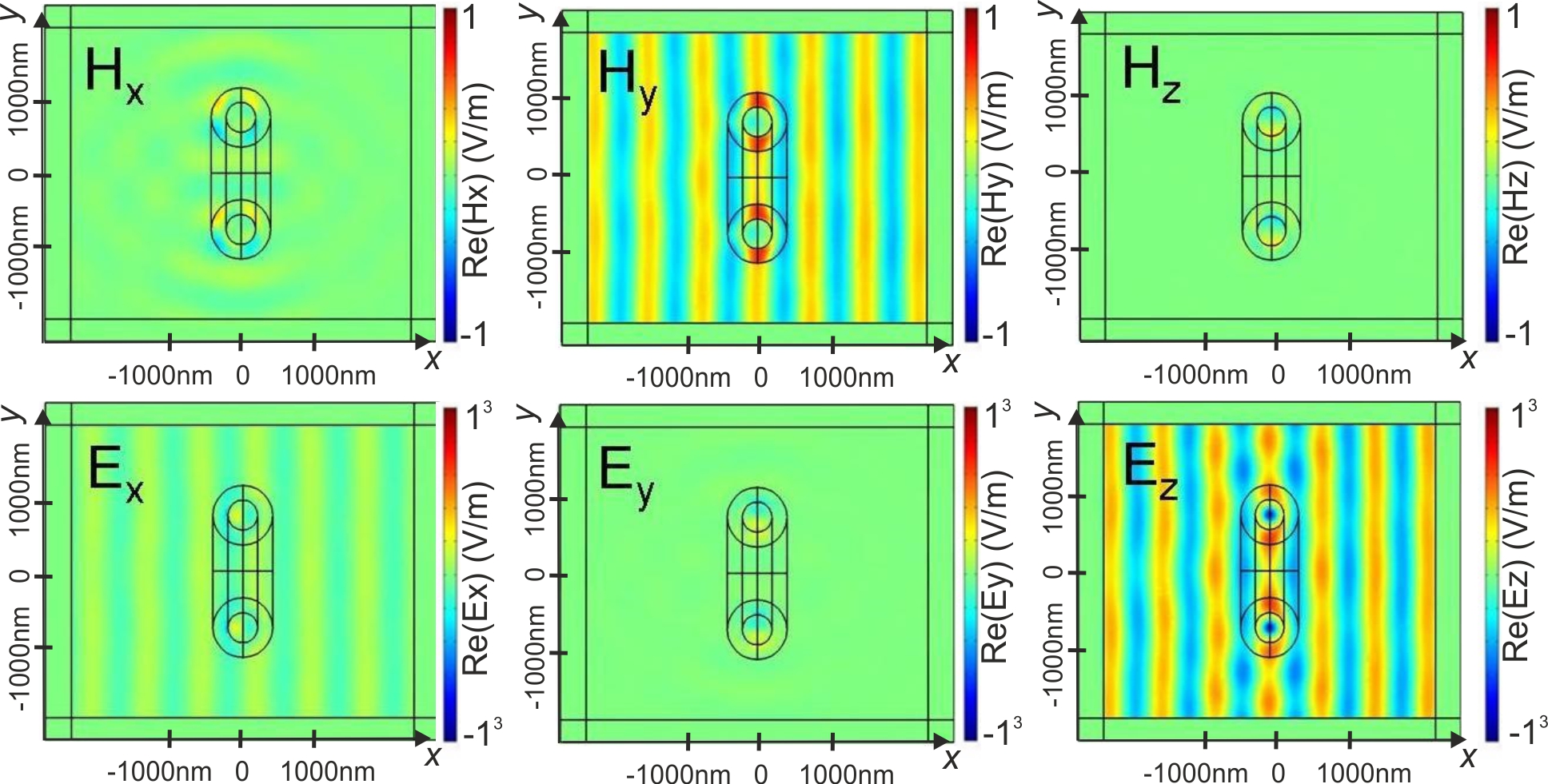}}
\vspace{0cm}
\caption{(Color online) All components of the electromagnetic field for the SH condition. Note the factor one thousand
between the color scales for the magnetic and electric field components.}
\label{fig17}
\end{figure}

\noindent Let us now describe the numerical results obtained using the three types of boundary conditions on the inner boundary of the wormhole (PMC, SH lining and transmission conditions). In figure \ref{fig11},
we implemented the PMC condition (upper panels) and the SH lining condition (lower panels) and further applied an electromagnetic field in the inner region). It can be seen that there are no significant
differences between the upper and lower panels, that is for the y-polarized magnetic field. From this, we can conclude that PMC conditions work well enough regarding the scattered magnetic field. However, if we now focus our attention on the (one non-vanishing) z-component of the electric field, cf. Figure \ref{fig16}, it transpires that SH lining conditions and PMC conditions lead to a dramatrically
different behaviour of the field inside the core of the wormhole: in the SH case, the magnitude of the field is large inside the core, while for PMC (and transmission) conditions it vanishes (as one would
expect for an invisibility cloak). Importantly, the electric field is non-singular across the interface between the coating and the core for both PMC and SH conditions, while the transmission condition
 leads to a singular field (indeed, the ${\bf T}$ matrix is singular at this interface. It is also illuminating to plot all the fields components. In Figure \ref{fig17}, one can see that the $H_y$ component
is indeed the dominant one, which is not surprising as we launched a y-polarised SPP. However, the other two comonents of the magnetic field do not vanish within the wormhole: indeed they couple
inside the spatially varying ansitropic coating: one should then note that the z-component of the electric field is stronger than the other two components, as it is essentially coming from the x and y-components of the magnetic field (through rotation by the {\it curl} operator). The fact that the Ez field is fairly large within the core of the wormhole is reminiscent of the almost trapped states in the
work of \cite{kuku}.

 \noindent From the mathematical viewpoint, the idea of Soft-Hard boundary conditions is very appealing \cite{sh} as it regularizes the electromagnetic field (thereby ensuring a well-posed problem in usual energy normed functional spaces). 
However, from the more pratical viewpoint, the PMC condition has the advantage of
preventing any field to penetrate the core region. Indeed, when an electromagntic mode propagates within the core of the wormhole, it would inevitably interfere with the exterior field which is non vanishing in case of SH lining conditions. Hence, the whole aspect of invisible tunnel would be spoiled.  
\section{Magnetic torus}
\begin{figure}[ht!]
\centerline{
\hspace{0cm}\includegraphics[width=15cm,angle=0]{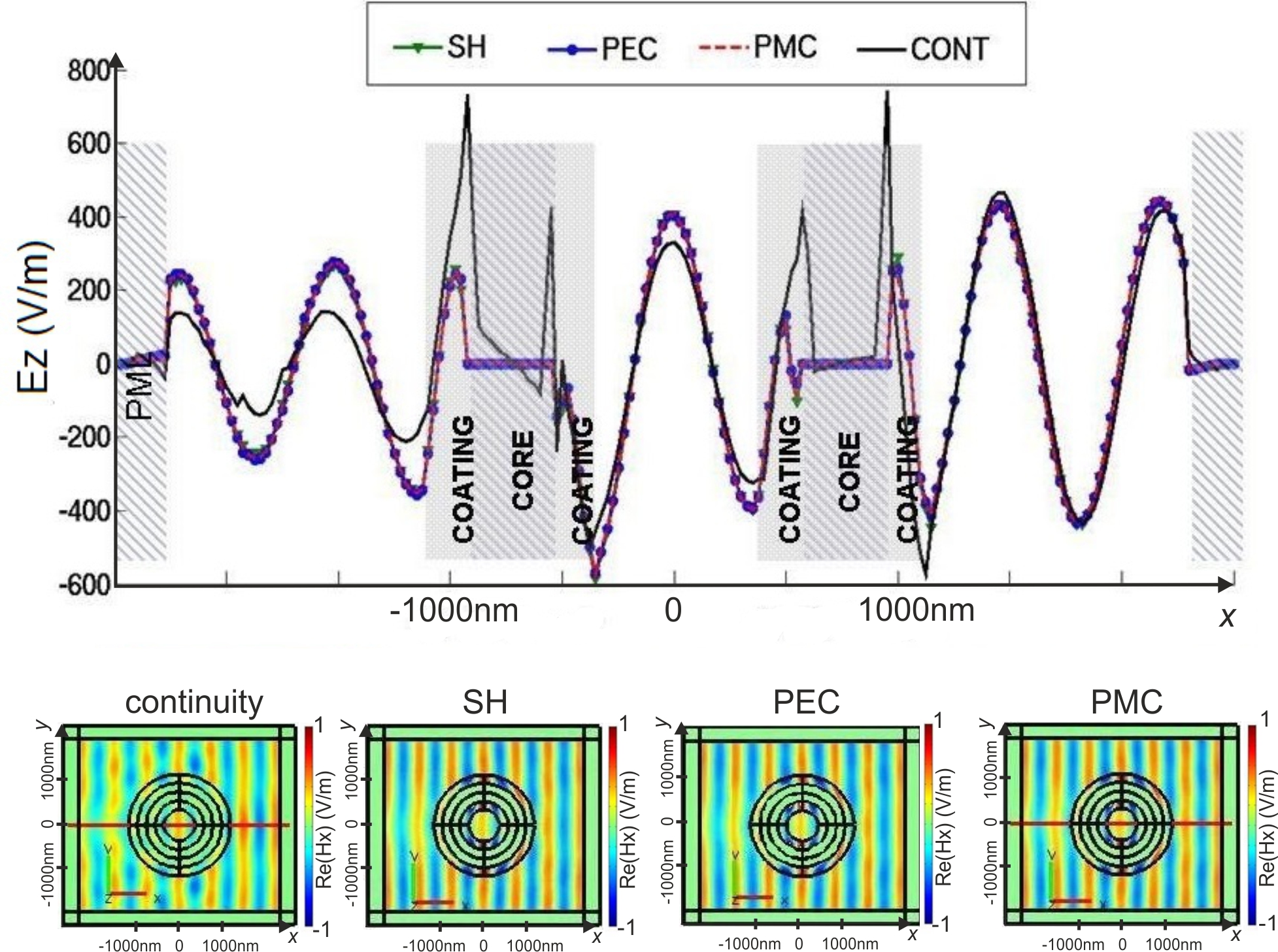}}
\vspace{0cm}
\caption{(Color online) Profile of $E_z$ in the plane $z=0$ along the red line shown in the inset, for three conditions on the inner boundary of the toroidal wormhole: Soft-Hard lining condition (green curve with triangles), continuity condition (continuous black curve), Perfect Electric conductor (PEC in dotted blue curve ) and Perfect magnetic Conductor (PMC in dashed red curve).
}
\label{fig18}
\end{figure}

\begin{figure}[ht!]
\centerline{
\hspace{0cm}\includegraphics[width=15cm,angle=0]{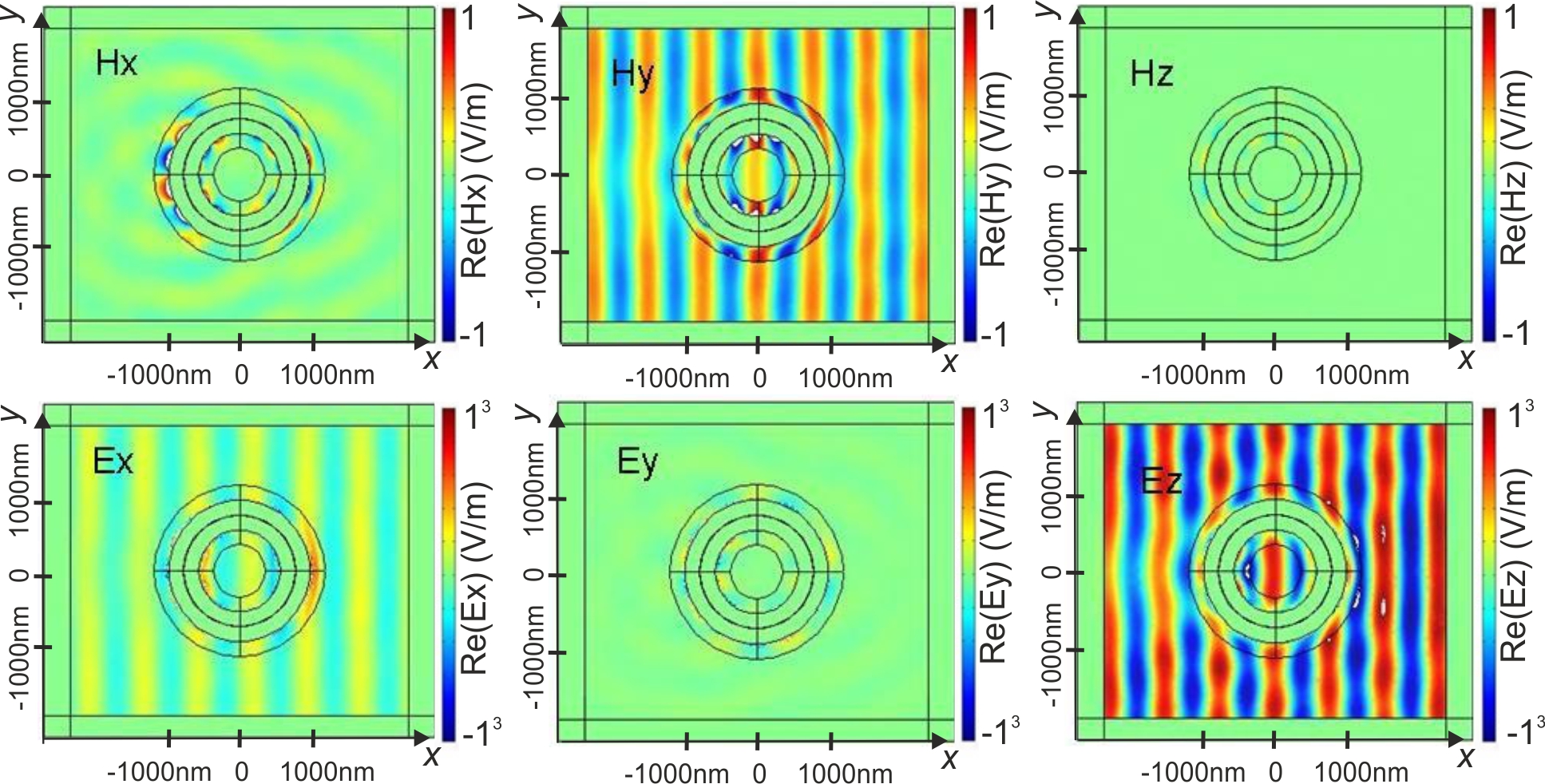}}
\vspace{0cm}
\caption{(Color online) All components of the electromagnetic field for the SH condition.
 Note the factor one thousand
between the color scales for the magnetic and electric field components.}
\label{fig19}
\end{figure}

A few words would be in order regarding the feasibility of the magnetic torus. Wood and Pendry have proposed in 2007 a route towards
magnetic metamaterials operating at low frequencies with a structure based on superconducting elements \cite{ben}. This theoretical
proposal has been since then experimentally validated \cite{gomory} with a magnetic cloak. The range of parameters required
for $\mu$ in Fig. \ref{fig8} is as for $\varepsilon$ in Fig. \ref{fig9}, and is therefore within experimental reach. We show in Fig. \ref{fig18} and \ref{fig19} the counterpart of Fig. \ref{fig16} and Fig. \ref{fig17} for the dielectric wormhole.

\section{Developed expression for the transformation matrix in Cartesian coordinates}

The toroidal coordinates $(r,v,w)$ can be expressed in terms of Cartesian coordinates $(x,y,z)$ as follows:
\begin{equation}
r=\frac{\sqrt{\rho(y^2\rho-2 y^2 R+z^2\rho-2 z^2 R+R^2\rho+x^2\rho)}}{\rho} \hbox{ , with $\rho=\sqrt{y^2+z^2}$,}
\end{equation}

\begin{equation}
v=\rm{Arctan}\left(\frac{y}{z}\right) \; , \; w =\rm{Arctan}\left( \frac{(\rho-R)\rho[\rho (y^2\sqrt {y^2+z^2}-2 y^2 R+z^2\rho-2 z^2 R+R^2\rho+x^2\rho)]^{-1/2}}
{x\rho \sqrt{\rho (y^2\rho-2 y^2 R+z^2\rho-2 z^2 R+R^2\rho+x^2\rho)}}\right).
\end{equation}
A systematic way to identify the tensors of permittivity and permeability associated with the toroidal geometry is actually to consider the Jacobian matrix $\bf{J}_{xr}$ associated with the opposite change of co-ordinate system. We emphasize the fact that it is the transformed domain and coordinate system $(r,v,w)$ that are mapped onto the initial domain with Cartesian coordinates $(x,y,z)$, and not the opposite:

\begin{center}
 \begin{equation}
  \left\{ \begin{array}{ccl}
          dx & = & \frac{\partial x}{\partial r}dr + \frac{\partial x}{\partial v}dv + \frac{\partial x}{\partial w}dw\\
        dy & = & \frac{\partial y}{\partial r}dr + \frac{\partial y}{\partial v}dv + \frac{\partial y}{\partial w}dw\\
        dz & = & \frac{\partial z}{\partial r}dr + \frac{\partial z}{\partial v}dv + \frac{\partial z}{\partial w}dw
\end{array} \right. \quad \Longleftrightarrow \quad
  \left( \begin{array}{c}
          dx\\
        dy\\
        dz
         \end{array} \right) = \bf{J}_{xr}
  \left( \begin{array}{c}
          dr\\
        dv\\
        dw
         \end{array} \right)
 \end{equation}
\end{center}
It follows that the transformation rule for expressing the tensors $\underline{\underline{\varepsilon'}}$
and $\underline{\underline{\mu'}}$ in the transformed coordinates in terms of the tensors $\underline{\underline{\varepsilon}}$
and $\underline{\underline{\mu}}$ in the Cartesian coordinates is \cite{andre}:
\begin{center}
 \begin{equation}
  \left\{ \begin{array}{ccl}
\underline{\underline{\varepsilon'}}(r,v,w) & = & \bf{J}_{xr}^{-1} \underline{\underline{\varepsilon}}(x,y,z)
\bf{J}_{xr}^{-T}{\mid{\rm det}(\bf{J}_{xr})\mid}\\
\underline{\underline{\mu'}}(r,v,w) & = & \bf{J}_{xr}^{-1} \underline{\underline{\mu}}(x,y,z)
\bf{J}_{xr}^{-T}{\mid{\rm det}(\bf{J}_{xr})\mid}\\
\end{array}\right.
 \end{equation}
\end{center}
where ${\rm det}(\bf{J}_{xr})$ is the determinant of the Jacobian matrix
and $\bf{J}_{xr}^{-T}$ denotes the inverse transpose matrix of $\bf{J}_{xr}$.

\noindent When the original permittivity and permeability matrices are proportional to the identity matrix,
which is our case, their transformed counterparts are given by:
\begin{equation}
{\underline{\underline{\varepsilon'}}}=\varepsilon {\bf T}_{xr}^{-1}
\; , \;
{\underline{\underline{\mu'}}}=\mu {\bf T}_{xr}^{-1}
\; , \;
\hbox{ where } {\bf T}_{xr}=\frac{{\bf J}_{xr}^T {\bf J}_{xr}}{\mid\rm{det}({\bf J}_{xr})\mid} \; .
\end{equation}

\noindent Thus far, we have only deduced the transformation matrix for transformed medium associated with
toroidal coordinates, but we did not take into account the stretch of toroidal coordinates $r'=a+r(b-a)/b$
used to blowup the centerline of the toroid in order to design the invisible handelbody of the wormhole.
It thus remains to compute the transformation matrix $\mathbf{T}$ as
expressed in the Cartesian co-ordinates $(x',y',z')$ associated with
the cylindrical cloak defined by the radii $R_1$ (interior radius)
the product of three elementary Jacobians :
\begin{equation}
\mathbf{J}_{xx'}=\mathbf{J}_{xr}\, \mathbf{J}_{rr'} \, \mathbf{J}_{r'x'} \; ,
\end{equation}
where $\mathbf{J}_{xr}$ (resp. $\mathbf{J}_{r'x'}$) is the Jacobian
associated with the change to toroidal coordinates ($r(x,y,z),v(x,y,z),w(x,y,z)$)
(resp. transformed Cartesian coordinates ($x'(r',v',w'),y'(r',v',w'),z'(r',v',w'$))
and where
\begin{equation}
\mathbf{J}_{rr'}=\mathrm{diag}(\frac{b}{b-a},1,1) \; ,
\end{equation}
is the radially contracted
toroidal coordinates, which as the same form as the one proposed in \cite{pendry}
for stretched polar coordinates in the context of cylindrical cloaks. Importantly, this
transform was used by You et al. \cite{you} for the design of toroidal cloaks in the
context of photonics, which leads to other kinds of metamaterials.

\noindent Finally, the material properties of the toroidal invisible handlebody bridges
the holes $D_1$ and $D_2$ on the metal surface are
described by the transformation matrix
\begin{equation}
{\underline{\underline{\varepsilon''}}}=\varepsilon {\bf T}^{-1}
\; , \;
{\underline{\underline{\mu''}}}=\mu {\bf T}^{-1}
\; , \;
\hbox{ where } \mathbf{T} := \mathbf{T}_{xx'} =
\frac{\mathbf{J}_{xx'}^T \mathbf{J}_{xx'}}{\mid\det(\mathbf{J}_{xx'})\mid} \; .
\end{equation}

Overall, the entries of the symmetric transformation matrix ${\bf T}$ are as follows:

\begin{equation}
T_{11}=-(-\beta^2 \cos(w)^2+2 r \cos(w)^2\beta-r^2) (R \alpha+r  \sin(w)- \sin(w)\beta)/(r (r-\beta)\alpha (R+r  \sin(w)))
\end{equation}

\begin{equation}
\begin{array}{llllll}
T_{12}=& \beta (-\beta+2 r) [-r  \sin(w) R+R  \sin(w)\beta+R  \sin(w) r\alpha-R^2\alpha+r^2-r\beta-r^2 \cos(w)^2+r \cos(w)^2\beta]  \\
& \times [\sin(v)  \sin(w) \cos(w)]/[r (r-\beta)\alpha (r^2 \cos(w)^2-r^2+R^2)]
\end{array}
\end{equation}

\begin{equation}
\begin{array}{llllll}
T_{13}=&\beta (-\beta+2 r) [-r  \sin(w) R+R  \sin(w)\beta+R  \sin(w) r\alpha-R^2\alpha+r^2-r\beta-r^2 \cos(w)^2+r \cos(w)^2\beta] \\
& \times [\cos(v)  \sin(w) \cos(w)]/[r (r-\beta)\alpha (r^2 \cos(w)^2-r^2+R^2)]
\end{array}
\end{equation}

\begin{equation}
\begin{array}{lllllll}
T_{22}=&-[(2 \cos(v)^2 \cos(w)^2 R  \sin(w)\alpha\beta^3+2 \cos(w)^2 \cos(v)^2 R^2\beta\alpha^2 r-6 R  \sin(w) \cos(v)^2\alpha r^2\beta \\
& \;\;\;\;\;\; +6 R  \sin(w) \cos(v)^2\alpha r\beta^2-2\alpha^2 R^2 r\beta \cos(w)^2+6 \cos(w)^2\beta^2 R  \sin(w) r\alpha \\
& \;\;\;\;\;\; +2 r^3 \cos(w)^4\beta+2 R  \sin(w)\alpha\beta^3-\beta^4+\cos(v)^2\beta^4+r^4 \cos(w)^2+4\beta^3 r-\alpha^2 R^2\beta^2 \\
& \;\;\;\;\;\; +2 R^2\beta\alpha^2 r-8 \cos(w)^2\beta^3 r-r^2\alpha^2 R^2-6 \cos(v)^2 \cos(w)^2 R  \sin(w)\alpha\beta^2 r \\
& \;\;\;\;\;\; -2 r^3 \cos(w)^4 \cos(v)^2\beta-2 r^3 R  \sin(w) \cos(v)^2+4 r^3 \cos(w)^2 \cos(v)^2\beta \\
& \;\;\;\;\;\; +\alpha^2 \cos(v)^2 R^2\beta^2-4 \cos(w)^4 \cos(v)^2\beta^3 r+8 \cos(w)^2 \cos(v)^2\beta^3 r \\
& \;\;\;\;\;\; +2 R^2 \cos(v)^2 r\beta-2 r^3 \cos(v)^2\beta-6 r^3 \cos(w)^2\beta-4 \cos(v)^2\beta^3 r+\cos(w)^4 \cos(v)^2\beta^4 \\
& \;\;\;\;\;\; -2 \cos(w)^2 \cos(v)^2\beta^4-R^2 \cos(v)^2\beta^2-\cos(w)^2 \cos(v)^2 R^2\beta^2\alpha^2-2\alpha^2 \cos(v)^2 R^2 r\beta \\
& \;\;\;\;\;\; -2 R  \sin(w) \cos(v)^2\alpha\beta^3-2 R  \sin(w) \cos(v)^2 r\beta^2-r^4-6 r\alpha\beta^2 R  \sin(w) \\
& \;\;\;\;\;\; -r^2 R^2 \cos(v)^2+r^2\alpha^2 \cos(v)^2 R^2+4 \cos(v)^2 r^2 \cos(w)^2 R  \sin(w)\alpha\beta \\
& \;\;\;\;\;\; -2 r^3  \sin(w) R\alpha+5 r^2 \cos(w)^4 \cos(v)^2\beta^2-10 r^2 \cos(w)^2 \cos(v)^2\beta^2 \\
& \;\;\;\;\;\; +5 r^2 \cos(v)^2\beta^2+11 r^2 \cos(w)^2\beta^2+2 r^3 R  \sin(w) \cos(v)^2\alpha+4 r^2 R  \sin(w) \cos(v)^2\beta \\
& \;\;\;\;\;\; +6 r^2 R  \sin(w)\alpha\beta+4 r \cos(w)^4\beta^3-6 r^2\beta^2+4 r^3\beta-2 R \cos(w)^2  \sin(w)\beta^3\alpha \\
& \;\;\;\;\;\;-5 r^2 \cos(w)^4\beta^2-4 r^2 R \cos(w)^2  \sin(w)\alpha\beta+\alpha^2 R^2\beta^2 \cos(w)^2-\cos(w)^4\beta^4+2 \cos(w)^2\beta^4)]\\

& \times 1/ [(r (r-\beta)\alpha (R^2\alpha+r^2-r\beta+r  \sin(w) R-R  \sin(w)\beta+R  \sin(w) r\alpha-r^2 \cos(w)^2+r \cos(w)^2\beta))]
\end{array}
\end{equation}

\begin{equation}
\begin{array}{llllll}
T_{23}&= \sin(v) \cos(v) \\
&\times [2\alpha^2 R^2 r\beta \cos(w)^2-6 \cos(w)^2\beta^2 R  \sin(w) r\alpha-2 r^3 \cos(w)^4\beta-2 R  \sin(w)\alpha\beta^3\\
& \;\;\;\;\;\;+\beta^4-4\beta^3 r+ \alpha^2 R^2\beta^2-2 R^2\beta\alpha^2 r+8 \cos(w)^2\beta^3 r+r^2\alpha^2 R^2+2 R^2 r\beta-2 r^3  \sin(w) R\\
& \;\;\;\;\;\;+4 r^3 \cos(w)^2\beta+6 r\alpha\beta^2 R  \sin(w)-r^2 R^2+2 r^3  \sin(w) R\alpha+4 r^2  \sin(w) R\beta-10 r^2 \cos(w)^2\beta^2\\
& \;\;\;\;\;\; -6 r^2 R  \sin(w)\alpha\beta-4 r \cos(w)^4\beta^3+5 r^2\beta^2-2 r R  \sin(w)\beta^2-R^2\beta^2-2 r^3\beta+2 R \cos(w)^2  \sin(w)\beta^3\alpha\\
& \;\;\;\;\;\;+5 r^2 \cos(w)^4\beta^2+4 r^2 R \cos(w)^2  \sin(w)\alpha\beta-\alpha^2 R^2\beta^2 \cos(w)^2+\cos(w)^4\beta^4-2 \cos(w)^2\beta^4] \\
&\times [1/{(r (r-\beta)\alpha (R^2\alpha+r^2-r\beta+r  \sin(w) R-R  \sin(w)\beta+R  \sin(w) r\alpha-r^2 \cos(w)^2+r \cos(w)^2\beta))}]
\end{array}
\end{equation}

\begin{equation}
\begin{array}{lll}
T_{33}=& [(2 \cos(v)^2 \cos(w)^2 R  \sin(w)\alpha\beta^3+2 \cos(w)^2 \cos(v)^2 R^2\beta\alpha^2 r-6 R  \sin(w) \cos(v)^2\alpha r^2\beta\\
& \;\;\;\;\;\;+6 R  \sin(w) \cos(v)^2\alpha r\beta^2+\cos(v)^2\beta^4-r^4 \cos(w)^2-6 \cos(v)^2 \cos(w)^2 R  \sin(w)\alpha\beta^2 r\\
& \;\;\;\;\;\;-2 r^3 \cos(w)^4 \cos(v)^2\beta-2 r^3 R  \sin(w) \cos(v)^2+4 r^3 \cos(w)^2 \cos(v)^2\beta+\alpha^2 \cos(v)^2 R^2\beta^2\\
& \;\;\;\;\;\;-4 \cos(w)^4 \cos(v)^2\beta^3 r+8 \cos(w)^2 \cos(v)^2\beta^3 r+2 R^2 \cos(v)^2 r\beta-2 R^2 r\beta+2 r^3  \sin(w) R\\
& \;\;\;\;\;\;-2 r^3 \cos(v)^2\beta+2 r^3 \cos(w)^2\beta-4 \cos(v)^2\beta^3 r+\cos(w)^4 \cos(v)^2\beta^4-2 \cos(w)^2 \cos(v)^2\beta^4\\
& \;\;\;\;\;\;-R^2 \cos(v)^2\beta^2-\cos(w)^2 \cos(v)^2 R^2\beta^2\alpha^2-2\alpha^2 \cos(v)^2 R^2 r\beta-2 R  \sin(w) \cos(v)^2\alpha\beta^3\\
& \;\;\;\;\;\;-2 R  \sin(w) \cos(v)^2 r\beta^2+r^4-r^2 R^2 \cos(v)^2+r^2 R^2+r^2\alpha^2 \cos(v)^2 R^2\\
& \;\;\;\;\;\;+4 \cos(v)^2 r^2 \cos(w)^2 R  \sin(w)\alpha\beta-4 r^2  \sin(w) R\beta+5 r^2 \cos(w)^4 \cos(v)^2\beta^2\\
& \;\;\;\;\;\;-10 r^2 \cos(w)^2 \cos(v)^2\beta^2+5 r^2 \cos(v)^2\beta^2-r^2 \cos(w)^2\beta^2+2 r^3 R  \sin(w) \cos(v)^2\alpha\\
& \;\;\;\;\;\; +4 r^2 R  \sin(w) \cos(v)^2\beta+r^2\beta^2+2 r R  \sin(w)\beta^2+R^2\beta^2-2 r^3\beta)] \\
& \times 1/ [(r (r-\beta)\alpha (R^2\alpha+r^2-r\beta+r  \sin(w) R-R  \sin(w)\beta+R  \sin(w) r\alpha-r^2 \cos(w)^2+r \cos(w)^2\beta))]
\end{array}
\end{equation}

where $R$ is the major radius and
\begin{equation}
\alpha=b-a \; , \; \beta=a \; .
\end{equation}

\end{document}